\documentclass{ws-ijseke}

\usepackage{amsmath,amssymb,amsfonts}
\usepackage{graphicx}
\usepackage{url}
\usepackage{booktabs}
\usepackage{xparse}
\usepackage{xspace}

\usepackage{xstring}
\def\ModSplit#1{\IfSubStr{#1}{|}{\StrBefore{#1}{|}[\before]\StrBehind{#1}{|}[\behind]\ModItem{\before}\-\ModSplit{\behind}}{\ModItem{#1}}}
\def\ModItem#1{\texttt{#1}}

\usepackage{framed}
\newcommand{\Conclusion}[1]{\begin{framed}\noindent #1\end{framed}}

\usepackage{cleveref}
\Crefname{figure}{Figure}{Figures}
\crefname{figure}{Fig.}{Figs.}
\crefname{table}{Table}{Tables}
\crefname{section}{Section}{Sections}

\newcommand{\qa}{\mathit{qa}}
\newcommand{\QG}{\mathit{Qual}}
\newcommand{\SCS}{\mathit{SCS}}
\newcommand{\RA}{\mathit{RA}}
\newcommand{\SC}{\mathit{Sem}}
\newcommand{\DS}{\mathit{DS}}
\newcommand{\VS}{\mathit{VS}}
\newcommand{\RAY}{\mathit{RA^+}}
\newcommand{\RAN}{\mathit{RA^-}}

\newcommand{\commits}{\mathit{commits}}

\newcommand{\Refactoring}[1]{\textsf{#1}}
\newcommand{\repository}[1]{\textit{#1}}
\newcommand{\Class}[1]{\ModSplit{#1}}
\newcommand{\Method}[1]{\ModSplit{#1}}
\newcommand{\Field}[1]{\ModSplit{#1}}

\newcommand{\dev}[1]{\textit{#1}}
\newcommand{\RQ}[1]{\textit{RQ}${}_{\mathrm{#1}}$}

\newcommand{\Time}{\mathit{time}}
\newcommand{\recency}{\mathit{recency}}

\newcommand{\Exp}{\mathrm{Exp}}
\newcommand{\rev}{\mathit{rev}}
\newcommand{\loc}{\mathit{loc}}
\newcommand{\pf}{pareto front\xspace}
\newcommand{\pfs}{pareto fronts\xspace}
\newcommand{\Var}[1]{\textit{#1}}
\newcommand{\VarN}[2]{$\mathit{#1}_{#2}$}

\begin{document}

\markboth{Chen and Hayashi}
{MORCoRA: Multi-Objective Refactoring Recommendation Considering Review Availability}

\title{MORCoRA: Multi-Objective Refactoring Recommendation\\Considering Review Availability}

\author{Lei Chen${}^*$ and Shinpei Hayashi${}^\dag$}

\address{School of Computing, Tokyo Institute of Technology\\
Ookayama 2--12--1, Meguro-ku, Tokyo 152--8550, Japan\\
${}^*$\email{chenlei@se.c.titech.ac.jp}
${}^\dag$\email{hayashi@c.titech.ac.jp}
}

\maketitle

\begin{abstract}\label{abstract}
\textit{Background:}
Search-based refactoring involves searching for a sequence of refactorings to achieve specific objectives. Although a typical objective is improving code quality, a different perspective is also required; the searched sequence must undergo review before being applied and may not be applied if the review fails or is postponed due to no proper reviewers.
\textit{Aim:}
Therefore, it is essential to ensure that the searched sequence of refactorings can be reviewed promptly by reviewers who meet two criteria: 1) having enough expertise and 2) being free of heavy workload. The two criteria are regarded as the review availability of the refactoring sequence.
\textit{Method:}
We propose MORCoRA, a multi-objective search-based technique that can search for code quality improvable, semantic preserved, and high review availability possessed refactoring sequences and corresponding proper reviewers.
\textit{Results:} 
We evaluate MORCoRA on six open-source repositories. The quantitative analysis reveals that MORCoRA can effectively recommend refactoring sequences that fit the requirements. The qualitative analysis demonstrates that the refactorings recommended by MORCoRA can enhance code quality and effectively address code smells. Furthermore, the recommended reviewers for those refactorings possess high expertise and are available to review.
\textit{Conclusions:} 
We recommend that refactoring recommenders consider both the impact on quality improvement and the developer resources required for review when recommending refactorings.
\end{abstract}


\section{Introduction}\label{s:introduction}
Refactoring is performed to improve the internal code structure without modifying its external behavior~\cite{fowler2018refactoring}.
Well-applied refactorings can considerably increase the maintainability, extendibility, and understandability of code.
However, determining the code that requires refactoring and specifying the type of refactoring to be applied demands significant expertise and time~\cite{how_we_refactor}.
Therefore, methods~\cite{kessentini2011design,ghannem2014model} have been proposed to automatically search for sequences of refactorings to optimize code quality through search techniques.
In addition, multi-objective optimization techniques have been adopted to search for refactorings that can both improve code quality and achieve other objectives such as preserving semantic coherence~\cite{ouni2012search} and eliminating code smells~\cite{ouni2017more}.

Similar to any other code changes, refactorings should be reviewed before their application~\cite{kagdi2008can,alomar2021refactoring}.
Kovalenko et al.~\cite{kovalenko2018does} also underscore the importance of selecting appropriate reviewers for refactoring recommenders, given that recommended refactorings often span multiple modules developed by different developers. 
However, selecting the appropriate reviewers for refactorings is a challenge~\cite{alomar2022code};
1) reviewers need to have expertise in the code where refactorings are applied on~\cite{thongtanunam2015should},
2) too many reviewers will cause unnecessary time consumption. As suggested by Rigby et al.~\cite{rigby2013convergent}, the reviewer count should be as low as fewer than three, and
3) the reviewers with high workloads should be avoided because they may not have the time to review~\cite{alomar2022code}.
We referred to achieving the above three requirements as possessing \emph{review availability}; refactorings reviewed by a small group of reviewers with the expertise and the time to review possess review availability.
Developers lacking expertise in the code where refactorings are applied are unsuitable to serve as reviewers for automated refactoring; they must first familiarize themselves with the source code, which is time-consuming.
When the reviewer count is high, significant time is required to confirm their availability to review, and communication among them can also be a time-consuming process~\cite{sadowski2018modern}.
If reviewers are burdened with a heavy workload, they may postpone or even ignore the review for the refactorings, which leads to the refactoring not being applied~\cite{rigby2008open}.

\begin{figure}[t]
  \centering
  \includegraphics[width=0.9\textwidth]{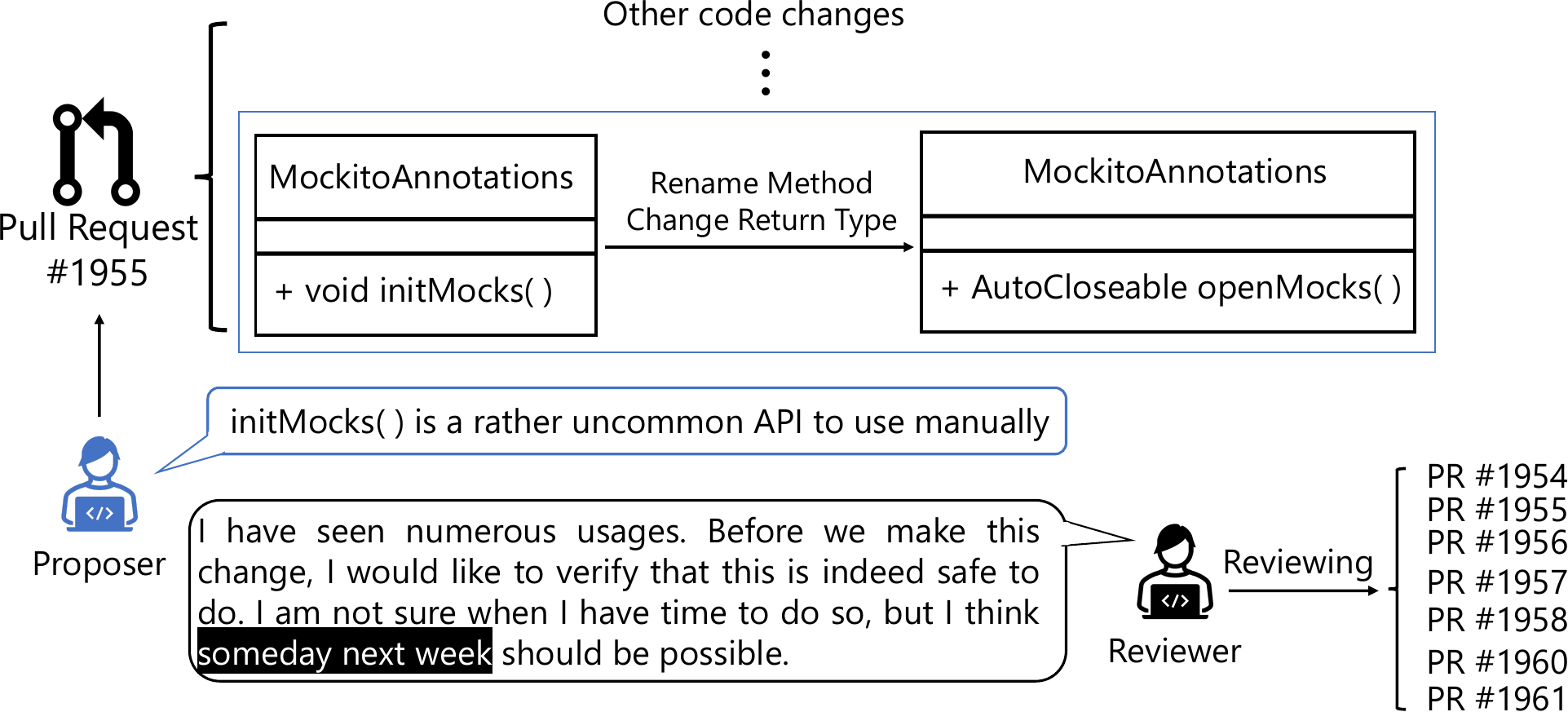}
  \caption{Motivating example of a pull request extract from \repository{mockito}.}
  \label{f:motivating_example}
\end{figure}

An example of a review of the refactorings being postponed because the reviewer has a high workload is found in an open-source repository~\repository{mockito} as displayed in \cref{f:motivating_example}.
This pull request\footnote{\url{https://github.com/mockito/mockito/pull/1955}} contains multiple refactorings and other non-refactoring code changes.
The refactorings include renaming the method from \Method{init|Mocks()} to \Method{open|Mocks()}, and also changing its return type from \Class{void} to \Class{Auto|Closeable}.
The proposer of the pull request commented that ``\Method{init|Mocks()} is a rather uncommon API to use manually''. He/she thought that it was fine to change its name and return type, and few users would be affected.
The reviewer refuted that the usage of the API was not unusual, and although a further investigation was required to review the refactorings, time constraints prevented him/her from conducting it, resulting in its postponement.
We examined the pull request history and found that the reviewer was reviewing the other seven pull requests from the time when this pull request was first proposed to the moment he/she decided to postpone the review of it.
The heavy workload of reviewing postponed the review of the refactorings.

If a search-based refactoring tool recommends the above refactorings during continuous integration, and there exists a short deadline and limited resources, which is a common scenario in practical development~\cite{rebaiMultiobjectiveCodeReviewer2020}, the refactorings are required to be reviewed and applied inside the limitation.
However, the heavy workload may cause the reviewer to give up reviewing, resulting in the refactorings not being applied.
This, in turn, leads to the decay of software quality~\cite{eick2001does}.
Therefore, a practical challenge in search-based refactoring is to search for refactorings that can improve code quality and possess high review availability.

To overcome this challenge, we propose MORCoRA, which is a search-based refactoring approach that recommends refactorings and corresponding reviewers based on a multi-objective evolutionary algorithm NSGA-II~\cite{Deb2002afast}.
We enhance the existing multi-objective search-based refactoring recommendation approach proposed by Ouni et al.~\cite{ouni2012search} in terms of the practicality of the technique by incorporating the review availability of recommended refactorings.
Refactorings recommended by our technique should achieve the following:
1) improving code quality,
2) preserving code semantic coherence after being applied, and
3) possessing high review availability.

The main contributions of this paper are as follows:
\begin{itemize}
    \item We proposed MORCoRA, a multi-objective search-based refactoring recommendation technique that considers review availability.
    \item We designed a metric named Review Availability to evaluate the review availability of refactoring.
    \item We evaluated MORCoRA on six open-source repositories. With quantitative and qualitative analysis, we conclude that MORCoRA can effectively recommend quality-improving, meaningful, and reviewable refactorings with proper reviewers.
    \item The performance of MORCoRA using different search algorithms on searching solutions is evaluated, and we conclude the NSGA-II to be the best.
    \item The MORCoRA is compared to the existing search-based refactoring technique~\cite{ouni2012search}, revealing superior performance in recommending reviewable refactorings that can improve the review availability by 433.8\%.
    In addition, we found that, instead of refactoring sequences with no proper reviewers, MORCoRA recommends refactorings that can improve code quality and be reviewed by highly expertise available reviewers.
\end{itemize}

The remainder of this paper is organized as follows.
\Cref{s:related work} outlines the related work.
The details of MORCoRA are introduced in \cref{s:methodology}.
We investigated three research questions in \cref{s:evaluation} to evaluate the effectiveness of MORCoRA and compared it with the existing tool.
The possible threats to validity is discussed in \cref{s:threats}.
Finally, we conclude this paper and describe directions for future work in \cref{s:conclusion}.

\section{Related Work}\label{s:related work}
\subsection{Search-Based Refactoring}
Search-based refactoring is applying search techniques to search for refactorings for the code.
For all search techniques, establishing clear objectives is imperative to provide guidance in their exploration.
The mathematical representations of the objectives are referred to as the fitness functions.
Based on the number of objectives, the studies of search-based refactoring can be categorized into two types: mono-objective and multiple-objective.

\subsubsection{Mono-Objective Approaches}
The mono-objective approach owns a single fitness function, which is typically composed of multiple metrics, to guide the search for the best sequence of refactorings.
Seng et al.~\cite{seng2006search} aggregated four code quality attributes, namely coupling, cohesion, complexity, and stability, into one fitness function to recommend the \Refactoring{Move Method} refactorings.
Code-Imp~\cite{o2008search} compared four search algorithms on searching a fitness function that contains three QMOOD~\cite{bansiya2002ahierarchical} quality attributes.
The findings indicated that two search techniques, namely hill climbing and simulated annealing, proved to be effective in search-based refactoring.

\subsubsection{Multi-Objective Approaches} \label{sss: related_work_multi-objective}
The second type considers multiple objectives when searching for refactorings.
In addition to the objective of improving code quality, 
different kinds of effort required for the searched refactorings are also important.

Morales et al.~\cite{morales2016finding} considered removing anti-patterns and reducing the test effort as the two objectives that the recommended refactorings should achieve.
They referred to the testing effort as the number of test cases required for each class and conducted experiments using four search algorithms.
The results revealed that the multi-objective cellular genetic algorithm~(MOCell)~\cite{nebroMOCellCellularGenetic2009} exhibited the best results, and their approach maintained a compromise between the number of anti-patterns corrected and the test cases required.

Typically, the recommended refactorings can improve code quality metrics, but these are not meaningful if code quality metrics are the only objective considered in search-based refactoring; the refactorings may violate domain semantics.
Ouni et al.~\cite{ouni2012search} considered preserving semantic coherence as the second objective, in addition to the code quality metrics, to make the recommended refactorings more meaningful.
The results showed that their approach could recommend more meaningful refactorings than the other approaches.
We also adopted this objective in this study.

To eliminate code smells, Ouni et al.~\cite{ouni2017more} proposed an approach that considers three objectives: 1) improving quality, 2) fixing code smell, and 3) introducing design patterns.
They used the NSGA-III algorithm~\cite{deb2013evolutionary} to search refactorings in seven open-source repositories.
Zhang et al.~\cite{zhang2024mirror}\ proposed MIRROR, which is a refactoring recommendation tool based on correlation analysis and multi-objective optimization. It considers three objectives: 1) improving quality, 2) removing code smell, and 3) maximizing the similarity to refactoring history and can recommend 26 types of refactorings.

Instead of using the code quality at the code implementation level as the main objective, Ivers et al.~\cite{ivers2022untangling} focused on the architecture level and proposed a technique to recommend refactorings to isolate entangled software.
They designed fitness functions for five objectives and conducted experiments on two of the objectives.
The genetic algorithm NSGA-II~\cite{Deb2002afast} was used as the search algorithm.
Their experimental results proved that the recommendations generated by their approach could reduce the problematic couplings that inhibited software isolation by more than 87\%.
In addition, developers were requested to evaluate their results, and 84\% of the recommendations were accepted.
Cortellessa et al.~\cite{cortellessa2023many}\ utilized NSGA-II to search for refactorings that can improve the performance and reliability of the software. They confirmed that their work can improve performance by up to 42\% and reliability by up to 32\%.

Our work belongs to the multi-objective approach, and we set three objectives to ensure that the searched refactorings are code quality improbable, meaningful, and reviewable.

\subsection{Reviewer Recommendation}
Assigning appropriate reviewers for code changes is critical because manually assigning is time-consuming and labor-intensive.
Studies have been conducted to address the reviewer assigning problem.

Based on the code change history, Balachandran~\cite{balachandran2013reducing} developed ReviewBot, a reviewer recommendation tool based on the change history of source code lines.
Instead of code changes, review history is also widely used to recommend reviewers.
Thongtanunam et al.~\cite{thongtanunam2015should} introduced REVFINDER to recommend reviewers based on the intuition that files located in similar file paths would be managed and reviewed by similarly experienced code reviewers.
They utilized the reviewed file path information in the review history to recommend reviewers.
Xia et al.~\cite{xia2015should} improved REVFINDER by proposing a hybrid approach based on text mining in review requests and the reviewed file path information to determine optimal reviewers.
Zanjani et al.~\cite{zanjani2015automatically} leveraged the number of review comments and their time spans in historical review data to recommend reviewers.

Code review is a human-involved process, and it has a social aspect that plays a crucial role in assigning reviewers~\cite{ouni2016search,kononenko2015investigating,bosu2013impact}.
Yu et al.~\cite{yu2014should} developed a comment network to recommend reviewers with the pull request history for a given repository. 
The comment network revealed the previous collaborations and relationships between reviewers.
They compared their approach with a machine learning-based approach, and the results revealed that their comment network method achieved considerable improvement.
Ouni et al.~\cite{ouni2016search} leveraged a search-based approach by considering two objectives: reviewer expertise and reviewer collaboration experiences.
Chouchen et al.~\cite{chouchen2021whoreview} improved the work of Ouni et al. by considering reviewer workload as the third objective.

The above works can only recommend reviewers to given code changes.
Our technique recommends both quality-improving code changes and their proper reviewers.

\section{MORCoRA Approach}\label{s:methodology}
\subsection{Overview}\label{ss: techniqe overview}
\begin{figure}[t]\centering
  \includegraphics[width=\textwidth]{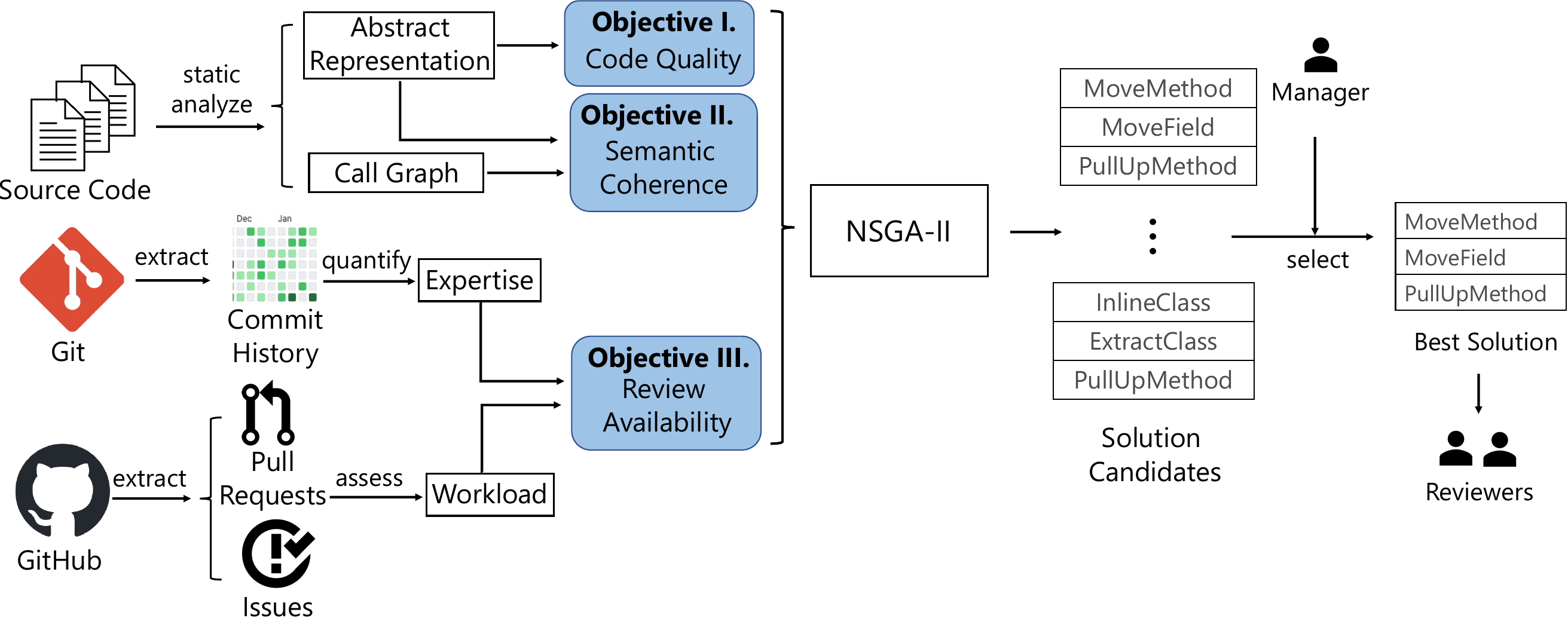}
  \caption{Process overview of MORCoRA approach.}
  \label{f:overview}
\end{figure}

The name of the proposed approach, MORCoRA, is derived from the key components of our technique and abbreviated from the title: \textbf{M}ulti-\textbf{O}bjective \textbf{R}efactoring Recommendation \textbf{Co}nsidering \textbf{R}eview \textbf{A}vailability.
In general, MORCoRA applies a multi-objective algorithm to search refactoring sequences that conform to three objectives: 1) improving code quality, 2) preserving semantic coherence, and 3) possessing high review availability.
We extended the study of Ouni et al.~\cite{ouni2012search} by considering the review availability objective.

An overview of MORCoRA is presented in \cref{f:overview}.
The inputs of MORCoRA are three-fold: 1) the target source code snapshot of the repository, 2) the commit history in the Git repository, and 3) the pull request history and issue history of the repository extracted from GitHub, where the repository is being hosted.
Static analysis is performed on the target source code snapshot to obtain its abstract representation and the call graph.
Abstract representation reflects the structural features of the source code, and a call graph records the invocation and access relations of the elements in the source code.
These two metrics evaluate whether the recommended refactoring sequences conform to the first two objectives.
The commit history is utilized to measure the expertise of contributors on files within the repository.
The pull request history and issue history are assessed to determine the workload of each contributor.
The resulting evaluations of expertise and workload are then used to determine whether each refactoring sequence aligns with the third objective.
The multi-objective evolutionary algorithm NSGA-II~\cite{Deb2002afast} searches for refactoring sequences that achieve the three objectives, i.e., gaining high scores in the corresponding fitness functions, which are used to quantify the three objectives.
The output of the NSGA-II is \textit{\pfs}, which are sets of candidate solutions, each comprising sequence of refactorings that provide a near-optimal trade-off among the three objectives.
Because the human-in-the-loop approach is suggested as the best method to help developers perform refactorings~\cite{simons2015search}, we let a manager of the repository choose the best solution among the candidates according to the project development needs.
The proper reviewers for the sequence can be deduced by MORCoRA.

\subsection{Static Analysis}\label{ss: static analysis}
A static analysis tool, JxPlatform2~\cite{jxplatform2}, is used to extract the abstract representation and the call graph from a target snapshot of the repository.

\textbf{Abstract Representation:} For each class in the source code, its name, inheritance relationship, member fields, and signatures of the methods are extracted as the abstract representation of this class.
Concrete implementations of methods are not used because only design-level refactoring types that modify the code structure are considered in this study.

\textbf{Call Graph:} For each class in the source code, its method invocation relations and member field accessing relations are extracted as a call graph.
The call graph is a directed graph, and its vertices represent the methods and fields of the classes, and the edges represent the invocation and access relations between vertices.
Following the Ouni et al.~\cite{ouni2012search}, only the static call graph is adopted and the dynamic bindings not resolved.

\subsection{Objective I: Improving Code Quality}\label{ss:objective i}
Based on Ouni et al.~\cite{ouni2012search} and also the established studies in the field of search-based refactoring~\cite{ouni2017more,koc2011empirical}, we used the Quality Model for Object-Oriented Design~(QMOOD)~\cite{bansiya2002ahierarchical} to evaluate the change in the code quality by performing recommended refactorings.
The QMOOD comprises six quality attributes\ of maintainability: reusability, flexibility, understandability, functionality, extendibility, and effectiveness.
In MORCoRA, these attributes are calculated using 11 low-level design metrics derived from the abstract representation of the source code.
We calculate the quality gain~($\QG$) caused by the recommended refactorings in the same manner as Ouni et al.~\cite{ouni2017more} as follows:
\begin{align}
  \QG_R = \sum_{i=1}^6 (\qa_i^R - \qa_i)
\end{align}
where $\qa_i$ and $\qa_i^R$ are the attributes before and after performing refactorings $R$, respectively.
Consequently, we transfer the objective of improving code quality to that of increasing $\QG$, which represents the improvement in the maintainability of the source code.

\subsection{Objective II: Preserving Semantic Coherence}\label{ss:objective ii}
To apply a recommended refactoring that affects multiple code elements, such as \Refactoring{Move Method}, ensuring that the source and target classes are semantically similar is essential to ensure the current refactoring is meaningful~\cite{ouni2012search}.
For example, moving a method \Method{study()} from the class \Class{Student} to the class \Class{Car} is not meaningful even though this refactoring may improve the value of quality metrics.
We follow the work of Ouni et al.~\cite{ouni2012search} and calculate the vocabulary-based similarity ($\VS$) and the dependency-based similarity~($\DS$).
More details are presented in the literature~\cite{ouni2012search}.

The semantic coherence score $\SCS$ between the two classes $c_1, c_2$ is calculated as follows:
\begin{align}
  \SCS(c_1,c_2) = \alpha \, \DS(c_1, c_2) +(1-\alpha) \, \VS(c_1,c_2)
\end{align}
where $\alpha \in [0,1]$ denotes a hyperparameter for blending the two types of similarities.
Semantic coherence $\SC$ for a sequence of refactorings $R$ is calculated as follows:
\begin{align}
  \SC_R = \frac{1}{|R|}\sum_{r \in R} \SCS(c_1^r, c_2^r)
\end{align}
where $c_1^r$ and $c_2^r$ denote the parameter classes for refactoring $r$.
We turn the objective of preserving semantic coherence into maximizing $\SC$.

\subsection{Objective III: Possessing High Review Availability}\label{ss:objective iii}
A refactoring sequence is considered to have high review availability if it can be reviewed by reviewers who possess expertise on the files where the refactorings will be implemented and the reviewers are not overwhelmed by heavy workloads.
Enough expertise enables reviewers to determine whether refactorings are proper and secure.
A low workload decreases the possibility of the review being postponed or ignored.
The details about the calculation of expertise and workload are introduced in the following sections.

\subsubsection{Reviewer Expertise}\label{ss:reviewer expertise}
We adopted and improved the expertise assessment from two works, Bird et al.~\cite{bird2011dont} and Ouni et al.~\cite{ouni2016search}, to achieve a higher accuracy expertise evaluation.
The code contributors' commit frequencies and recency to a file in the commit history are used to evaluate their expertise on that file.

In Bird et al.~\cite{bird2011dont}, the expertise of a developer on a file is calculated using his/her commit frequency, which is the ratio of the number of commits made by the developer to that file and the total number of commits in that file.
The commit frequency indicates a developer's knowledge of a file.
However, the recency of the commits is not considered in this definition.
The expertise of a developer who only proposed commits a long time ago should be lower than that of a developer who proposed commits more recently, even if the numbers of proposed commits by these two developers are the same.
The code may undergo considerable changes during development, and the previous developer may have been no longer familiar with it.

Instead of code commits, Ouni et al.~\cite{ouni2016search} use developers' previous review comment histories on a code review platform to evaluate their expertise.
Both the frequency and recency of comments are considered for higher accuracy expertise calculation.
When it comes to open-source repositories on a code hosting platform, such as GitHub, the comments are typically found in pull requests and issues.
Those comments contain many non-review-related content, such as discussions and inquiries.
As a result, they are not considered as the direct indicators of developers' knowledge of the source code, unlike commits.
The idea of using recency is applicable to commits, and a study proved that developers contributing in the more recent in a higher frequency should have higher expertise~\cite{chouchen2021whoreview}.

To this end, we present the expertise of open-source repository developers based on the commit frequency and recency.

We calculate the frequency the same way as introduced by Bird et al.
We improve the method by Ouni et al. to calculate recency more accurately.
In the original method, recency values for comments added at different time points in a pull request are not distinguished; they have the same recency value, which is calculated by using the added time of the most recent comment.
However, in the case of code commits, the expertise of a developer who made three commits many years ago and one commit recently should be lower than that of another developer who made four commits recently; frequent recent commits reveal that the developer's current interest in the target code is high.
A different approach was taken to prevent the above scenario, where a single recent commit would boost the recency value of all existing commits.
For every commit, we computed a value that reflects its relative recency.
The recency for a commit $c$ proposed at time $\Time(c)$ is calculated as follows:
\begin{align}
    \recency(c) = \frac{\Time(c)-\Time(c_0)}{\Time(c_*)-\Time(c_0)}
\end{align}
where $c_0$ and $c_*$ are the most initial and the most latest commit of the repository, respectively.
In this way, the weight of the previously proposed commit can be deduced.

We used recency as the weight of a commit to emphasize more recent commits and calculate the expertise of a developer $d$ for a file $f$ as follows:
\begin{align}
    \Exp(d,f) = \min \left\{1, \frac{\sum_{c \in \commits_d(f)}{\recency(c)}}{\min\{|\commits(f)|, T_c\}}\right\}
\end{align}
where $c$ is a commit, $\commits(f)$ is the set of commits in the file $f$, and $\commits_d(f)$ is the set of commits proposed by $d$ in the file $f$.
The $T_c$ is a positive integer hyperparameter.
By utilizing the sum of the recency of commits proposed by a developer in the file to divide the total number of commits, we normalize the expertise value.
We introduced two workarounds into this formula.
First, to avoid some developers' recent significant contributions to files having a very long commit history not being appreciated well, we set an upper bound $T_c$ for the total number of commits to be used as a denominator.
Second, to avoid the unusually high expertise of developers who contributed most of the commits in history, we set the maximum value of the formula to 1.

\subsubsection{Workload}\label{ss:workload}
MORCoRA takes into account two types of workload: \textit{communication workload} and \textit{in-progress contribution workload}.
In terms of the \textit{communication workload}, gathering a group of reviewers who possess sufficient expertise to evaluate a refactoring sequence that spans various modules may result in the group size being large.
The effort to find such a group is non-trivial~\cite{sadowski2018modern}, and the communication in the later review is also costful~\cite{bosu2013impact}.
The \textit{in-progress contribution workload} refers to the pull requests and issues the developers are currently working on.
As shown in the motivating example in \cref{s:introduction}, the review of refactoring may be postponed if the reviewers' in-progress workload is high.

Limiting the required maximum number of reviewers can effectively restrict the communication workload.
As suggested by Rigby et al.~\cite{rigby2013convergent} that the optimal number of reviewers in OSS is two, and one reviewer is also enough in some cases~\cite{sadowski2018modern}, we limited the maximum number of reviewers for a refactoring sequence to two in this paper.

The total number of pull requests and issues a developer $d$ participates in a certain time period is used as his/her in-progress contribution workload and represented as follows:
\begin{align}
    \mathrm{WL}(d) = |\mathit{PR}_{t_0,t_e}(d)|+|\mathit{IU}_{t_0,t_e}(d)|
\end{align}
where hyperparameters $t_0$ and $t_e$ are the start and end time points of the time period.
$\mathit{PR}_{t_0,t_e}(d)$ is the set of pull requests that were proposed or received comments by $d$ in the period from time $t_0$ to $t_e$.
Similarly, $\mathit{IU}_{t_0,t_e}(d)$ is the set of issues that were proposed or received comments by $d$ in the period from time $t_0$ to $t_e$.

\subsubsection{Review Availability}\label{ss:review availability}
The review availability of a sequence of refactoring $R$ is presented below:
\begin{align}
    \RA_R &= \frac{1}{|R|}\sum_{d \in \rev(R)}\sum_{f \in \loc(R)} \Exp(d,f), \\
   \rev(R) &= \{d \mid \forall f \in \loc(R), \mathrm{WL}(d) \leq T_w \wedge \Exp(d,f)>0\}
\end{align}
where $loc(R)$ indicates a multiset of files where any refactoring in $R$ is applied on.
The $T_w$ is a hyperparameter set for the workload that any reviewer having a workload larger than it is regarded as unable to review.

In this way, our objective of achieving high review availability is transformed into the goal of maximizing $\RA$.

\subsection{NSGA-II Adoption for Search-Based Refactoring}\label{ss:multi-objective search algorithm adoption for search-based refactoring}
This subsection introduces how we adopted the NSGA-II algorithm to search for sequences of refactorings that fit our three objectives.

NSGA-II is one of the multi-objective evolutionary search algorithms (MOEAs) that is widely used in the field of search-based software engineering.
It is inspired by the process of natural selection, where the fittest individuals are selected to reproduce the offspring of the next generation~\cite{Deb2002afast}. 
Specifically, the algorithms comprise four main steps:
1) solution representation, 2) crossover, 3) mutation, and 4) selection.

\begin{figure}[t]\centering
  \includegraphics[width = 6cm]{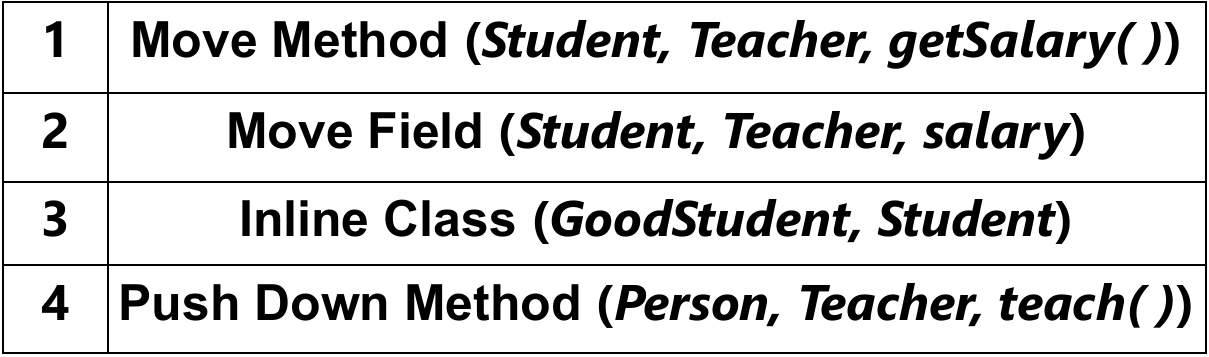}
  \caption{Example of a solution representation.}
  \label{f:solution example}
\end{figure}
\begin{table}[tb]\centering
  \tbl{Refactoring Operations\label{t:refactoring operations}}
  {\begin{tabular}{@{}cl@{}} \toprule
        Index & Refactorings \\ \midrule
        1 & \Refactoring{Inline Class} (\VarN{class}{1}, \VarN{class}{2}) \\ 
        2 & \Refactoring{Move Method} (\VarN{class}{1}, \VarN{class}{2}, \Var{method})\\
        3 & \Refactoring{Pull Up Method} (\VarN{class}{1}, \VarN{class}{2}, \Var{method})\\
        4 & \Refactoring{Push Down Method} (\VarN{class}{1}, \VarN{class}{2}, \Var{method})\\
        5 & \Refactoring{Move Field} (\VarN{class}{1}, \VarN{class}{2}, \Var{field})\\
        6 & \Refactoring{Pull Up Field} (\VarN{class}{1}, \VarN{class}{2}, \Var{field})\\
        7 & \Refactoring{Push Down Field} (\VarN{class}{1}, \VarN{class}{2}, \Var{field})\\
        8 & \Refactoring{Null} \\\botrule
  \end{tabular}}
\end{table}

\subsubsection{Solution Representation}
A sequence of refactoring operations is called one solution.
The first step of the search algorithm involves describing this solution.
As shown in \cref{f:solution example}, we describe the solution as a vector, and each of its dimensions is assigned a refactoring operation.
The position of each refactoring operation determines the order in which it will be performed.
Refactoring operation types considered in this paper are listed in \cref{t:refactoring operations}.
Each row is one refactoring operation and its parameters.
For example, \Refactoring{Move Method}(\textit{class1}, \textit{class2}, \textit{method}) represents the moving of the \textit{method} from \textit{class1} to \textit{class2}.
The last type \Refactoring{Null} means that this refactoring operation does nothing.
Since the maximum number of the solution's dimensions must be set manually, if the optimal number of refactoring operations required by a specific repository is smaller than the manually set maximum length, the \Refactoring{Null} operation will fill the blank.
This way, the number of refactoring operations can be tuned by the search algorithm according to different repositories without acknowledging the optimal number of refactoring operations required by those repositories. 

\subsubsection{Crossover}
\begin{figure}[t]\centering
      \includegraphics[width=0.8\textwidth]{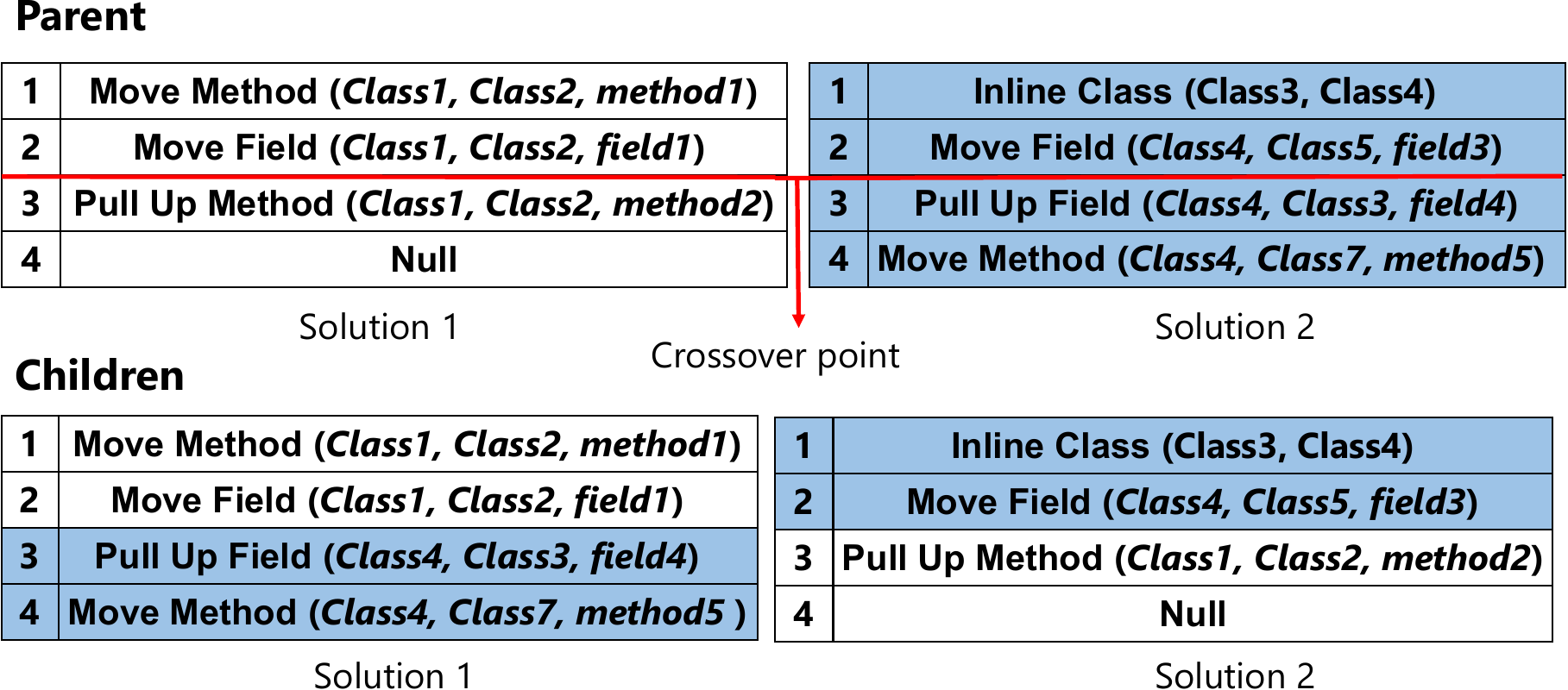}
      \caption{Example of crossover.}
      \label{f:crossover}
\end{figure}

Crossover is the process of combining the genetic information of two parent solutions to generate new offspring.
We use the one-point crossover operator, simulated binary crossover (SBX crossover)~\cite{deb1995simulated} to perform crossover.
SBX crossover can ensure that the children solutions are close to the parents before and after crossover, which is essential to perform an effective search in multi-objective problems~\cite{jain2013evolutionary}.
As depicted in~\cref{f:crossover}, the first step is to find a point on the parent solutions to divide each into two parts. 
Then the first offspring is generated by combining the first part of the first parent and the second part of the second parent, and vice versa for another offspring.

\subsubsection{Mutation}
\begin{figure}[t]\centering
      \includegraphics[width=5cm]{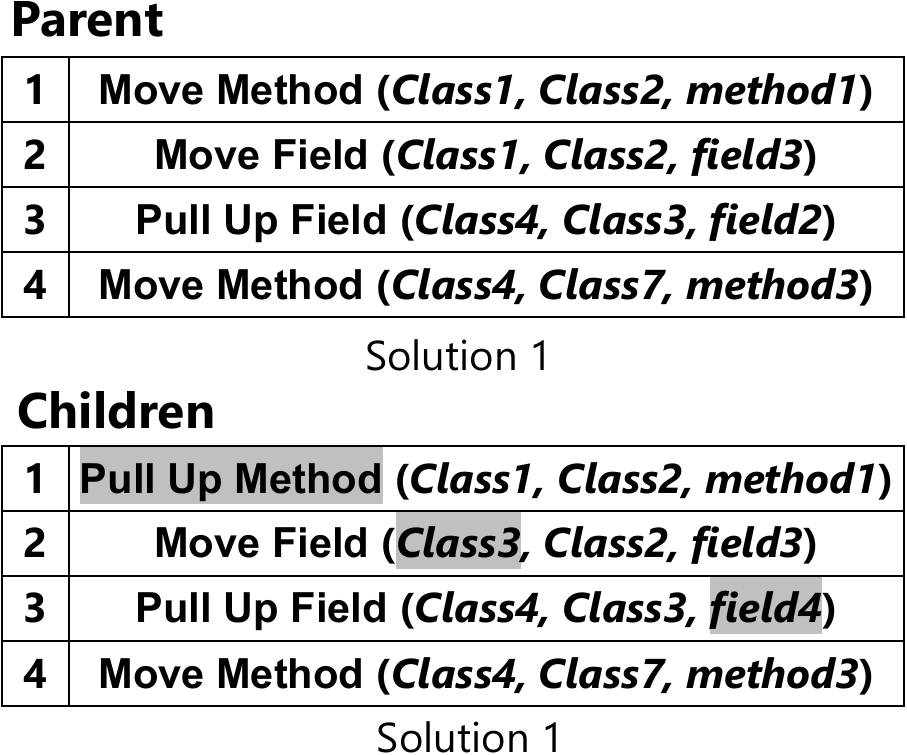}
      \caption{Example of mutation.}
      \label{f:mutation}
\end{figure}
Mutation is to randomly change elements in the solution to prevent the search from getting stuck at a local optimal result.
An example of mutation is shown in \cref{f:mutation}, the refactoring operation type of the first refactoring is changed to \Refactoring{Pull Up Method}, and the source class in the parameter of the second refactoring operation is changed to \Class{Class3}.
In the third refactoring operation, the pulled-up field is mutated to \Field{field4}, and there is no change in the fourth refactoring operation.
Mutation may introduce invalid refactorings, and those refactorings are regarded as \Refactoring{Null}.

\subsubsection{Selection}
Selection is to evaluate and select optimal solutions from the candidates.
In the evaluation, fitness functions are designed to guide the search.
The fitness functions in this study are the introduced metrics $\QG$, $\SC$, and $\RA$.
For each of the candidate solutions, the metrics will be calculated. 

NSGA-II uses fast non-dominated sorting to rank the solutions with the calculated metrics into different \pfs.
One solution is considered to dominate another if it has at least the same values in two metrics and a strictly higher value in another metric.
Solutions with the same domination count, which is the number of solutions that dominate the current solution, are assigned the same \pf.
The solutions in the same \pf are non-dominated with each other.
Among the \pfs, the first front contains non-dominated solutions, and the second front contains solutions dominated only by solutions in the first front, and so on.

Within each front, the crowding distance metric, which measures how close a solution is to its neighbors in the objective space, is calculated for each solution.
Solutions with a larger crowding distance are preferred because they contribute more to the diversity of the population, which facilitates the solution search.

Solutions are then selected for the next generation based on their front ranking and crowding distance.
The algorithm aims to maintain a diverse set of non-dominated solutions.
The selection process involves iteratively choosing individuals from the fronts until the next generation is filled.

\def\RQone{To what extent can MORCoRA improve code quality with meaningful reviewable refactorings?}
\def\RQtwo{How does MORCoRA perform on searching for refactorings using different evolutionary algorithms?}
\def\RQthree{How does MORCoRA compare to the existing search-based refactoring technique?}

\section{Evaluation}\label{s:evaluation}
This section details the experiments that we conducted to evaluate the effectiveness of MORCoRA.
We stated the following three research questions (RQs) that we addressed:
\begin{description}
  \item\textbf{\RQ{1}}: \RQone
  \item\textbf{\RQ{2}}: \RQtwo
  \item\textbf{\RQ{3}}: \RQthree
\end{description}

The experiment was conducted on a dataset containing six open-source Java repositories.
To evaluate \RQ{1}, we applied MORCoRA on the dataset and analyze the results from both quantitative and qualitative aspects.
To answer \RQ{2}, we compared the results of MORCoRA using five different multi-objective searching algorithms.
In \RQ{3}, the performance of MORCoRA was compared with the search-based refactoring technique benchmark proposed by Ouni et al.~\cite{ouni2012search}, which does not consider the review availability objective.
The detailed motivations for proposing those RQs will also be explained in the following subsections.

\subsection{Dataset}\label{ss:dataset}
Experiments were conducted on high-popularity and active Java repositories selected from GitHub.
We used the search engine of GitHub to collect Java repositories that meet the following criteria:
1) the popularity of a repository was evaluated based on the number of stars it has, and we only selected repositories with at least 10,000 stars, 
2) to ensure that the repositories were still active, we only considered those with recent commits, pull requests, or issues within half a year before April 2023, and
3) the repository size was evaluated by the number of classes contained, we selected repositories containing at least 200 and at most 1,500 non-test classes for our dataset.
Too few classes would result in a simple structure, making an exhaustive search more effective than an evolutionary algorithm, while too many classes would require a considerable amount of computation time.
Details about the selected repositories are provided in \cref{t:dataset_2}.

\begin{table}
  \centering
  \tbl{Dataset\label{t:dataset_2}}
    {\tabcolsep=6.5pt\begin{tabular}{@{}lrrrccr@{}}
          \toprule
          Repository &\# classes &\# commits &MEN &$t_0$& $t_e$ & \# PRevs\\
          \midrule
            \repository{auto} & 225 & 1,519 & 22,500 & 2022--11--26& 2022--12--02 & 70\\
            \repository{lottie-android} & 290 & 1,499 & 29,000 & 2022--03--08& 2022--03--14 & 88\\
            \repository{disruptor} & 440 & 1,306 & 44,000 & 2023--02--17& 2023--02--23 & 41\\
            \repository{mockito} & 513 & 5,822 & 51,300 & 2023--03--12& 2023--03--18 & 226\\
            \repository{retrofit} & 1,165 & 2,028 & 116,500 & 2022--12--09& 2022--12--15 & 95\\
            \repository{glide} & 1,428 & 2,832 & 142,800 & 2023--03--02& 2023--03--08 & 135\\
        \botrule
    \end{tabular}}
\end{table}

\subsection{Hyperparameter Settings}\label{ss:hyperparameter settings}
The similarity blending $\alpha$~(\cref{ss:objective ii}) was set to 0.8 because $\DS$ is more effective for preserving semantic coherence than $\VS$.
The $T_c$~(\cref{ss:reviewer expertise}) was set to 10 to prevent the candidate reviewers from being concealed in the long commit history file.

For calculating the workload in \cref{ss:workload}, following the reviewer recommendation work proposed by Rebai et al.~\cite{rebaiMultiobjectiveCodeReviewer2020}, the time period length between $t_0$ and $t_e$ was set to seven days. 
The criteria for determining the time period length are: 1) to avoid a too-long time period because work may have been finished and could not be counted as workload, and 2) to avoid a too-short time period where some unfinished work is omitted to be counted as workload.
Next, we picked one of the dates that a pull request or issue was updated within half a year before April 2023 as the end time $t_e$ to ensure that some developers have workload in reviewing; in other words, not all developers are exempt from the workload, which is not a scenario MORCoRA aims to address.
Each repository's specific value of $t_e$ is presented in \cref{t:dataset_2}.
The parameter $T_w = 2$ was used to simulate a situation to avoid asking busy developers to review.
Based on the $t_w$, $t_e$, and $T_w$, the number of possible reviewers (PRevs), those with enough expertise to review and are free of heavy review workload, are calculated and listed in \cref{t:dataset_2}.

The parameters used for the MOEAs are consistent with what is recommended in published guidelines\footnote{\url{https://www.obitko.com/tutorials/genetic-algorithms/recommendations.php}}.
Specifically, we set the crossover probability to 0.90 and the mutation probability to 0.05.
The population size in the algorithms was set to 100; each population contained 100 refactoring sequences.
Because the selected repositories were well-maintained and high quality, they may not require many refactorings.
Consequently, we set each solution to include a maximum of five refactoring operations. 
The algorithms determined the specific number for each repository by assigning some of the operations as \Refactoring{Null}.
The \textit{maximum evaluation number}~(MEN) was set to determine when to stop the search.
As larger repositories with more classes have larger search spaces and require more iteration of evaluations, different MEN values were set for each repository based on its size.
The larger repositories have higher MEN, and the concrete numbers are shown in \cref{t:dataset_2}.

\subsection{\RQ{1}: \RQone} \label{ss:rq1}
\subsubsection{Motivation and Study Design}\label{sss:rq1_motivation}
The aim of our study is to determine the effectiveness of MORCoRA in searching refactoring sequences that conform to three objectives so that it is regarded as useful from the perspective of software maintainers.
We applied MORCoRA on our dataset with the settings introduced in \cref{ss:hyperparameter settings} to 
investigate the above problem.
The output of MORCoRA on each repository was a \pf.
We performed five repeated runs because of the stochastic nature of the algorithms.
Because the solutions with negative $\QG$ violate the purpose of refactorings, only solutions with non-negative $\QG$ were selected from the \pfs for analysis. 
In addition, 60 solutions were randomly selected to provide a reasonably indicative sample, and one of the authors manually assessed refactoring sequences to determine whether they conform to the three objectives. 
The author responsible for the assessment has extensive experience in software refactoring and code review, ensuring a thorough and knowledgeable analysis.
A signal of code quality improvement~\cite{fowler2018refactoring} is that code smells are eliminated in addition to the improvement in code quality metrics.
In this work, the widely-used code smell detection tool Jdeodorant~\cite{tsantalis2017clone} was used to detect the code smells in the dataset. 
One of the authors manually assessed 60 solutions, which were randomly selected from the \pfs, to determine whether the searched refactoring sequences could eliminate code smells and were also meaningful and reviewable.
We checked the refactoring diff to determine the meaningfulness of the refactoring sequences. 
We inspected the history of developers' activities to guess the situation when the pull request including the searched refactoring sequences was proposed to determine whether the pull request is reviewable or not.
As a first pilot study of MORCoRA, we decided not to conduct a human study to ask developers of repositories in our dataset to review the searched refactoring sequences; it may be a costful work for them.
It is future work to conduct such a human study.

\subsubsection{Results and Discussions}\label{sss:rq1_results}
The result of the performance of MORCoRA on each repository is shown in \cref{t:morcora_performance}.
The \textit{Avg.} and \textit{Med.} represent average and median values for each objective, respectively, and the bold values indicate the highest values in average/median across all repositories.
\begin{table}
  \centering
  \tbl{MORCoRA performance \label{t:morcora_performance}}
  {\begin{tabular}{@{}lccccccc@{}}
      \toprule
        & \multicolumn{2}{c}{$\QG$} & \multicolumn{2}{c}{$\SC$} & \multicolumn{2}{c}{$\RA$} \\ \cmidrule(lr){2-3} \cmidrule(lr){4-5} \cmidrule(lr){6-7}
      Repository & Avg. & Med. & Avg. & Med. & Avg. & Med. \\ \midrule
      \repository{auto} & 0.004 & 0.004 & 0.074 & 0.070 & 0.494 & 0.392 \\
      \repository{lottie-android} & 0.008 & 0.004 & 0.301 & 0.217 & 0.280 & 0.230 \\
      \repository{disruptor} & \textbf{0.009} & \textbf{0.008} & 0.076 & 0.063 & \textbf{1.678} & \textbf{1.570} \\
      \repository{mockito} & 0.003 & 0.003 & 0.082 & 0.077 & 1.495 & 1.235 \\
      \repository{retrofit} & 0.008 & 0.008 & \textbf{0.345} & \textbf{0.276} & 1.480 & 1.485 \\
      \repository{glide} & 0.003 & 0.003 & 0.074 & 0.064 & 0.243 & 0.171 \\ \botrule
  \end{tabular}}
\end{table}

In the repository \repository{disruptor}, both the highest average values for $\QG$ and $\RA$ as well as the median values for $\QG$ and $\RA$ are found.
For $\SC$, the highest average and median values are found in the repository \repository{retrofit}.
The maximum value for $\QG$ was 0.029, the $\SC$ ranged from 0.000 to 0.830, and the $\RA$ ranged from 0.000 to 4.000.
The values of the solutions for the three objectives were above zero, meaning that the refactoring sequences searched by MORCoRA could achieve the three objectives.

One of the authors manually assessed and analyzed 60 randomly selected solutions.
The result showed that 35 out of 60 solutions were meaningful, reviewable, and eliminated at least one code smell, indicating that MORCoRA achieves a precision of 58.3\%.
Note that we did not calculate the recall metric because there is no standard solution dataset that provides the best refactoring sequences.

\begin{figure}[tb]
  \centering
  \includegraphics[width=0.7\textwidth]{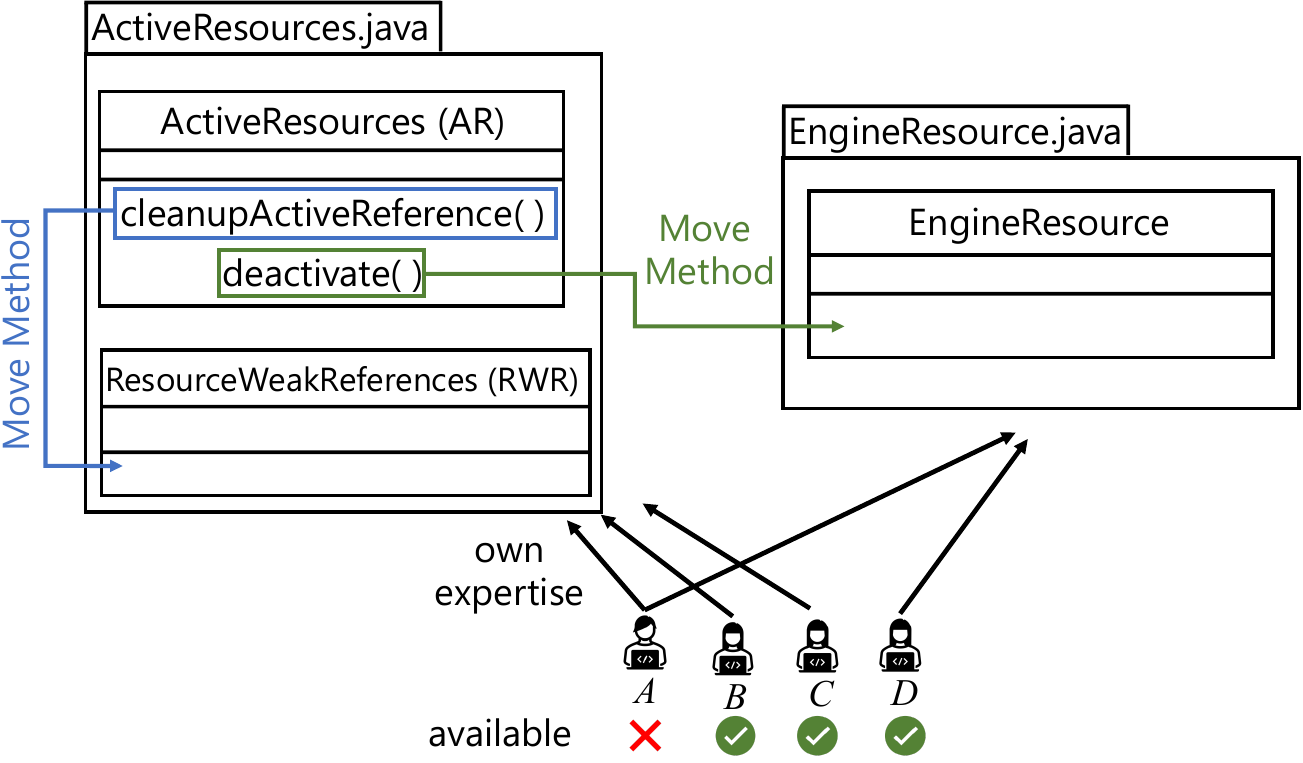}
  \caption{Example solution recommended by MORCoRA.}
  \label{f:positive_example}
\end{figure}

An example of such a solution is shown in \cref{f:positive_example}.
In the repository $\repository{glide}$, the method \Method{cleanup|Active|Reference()} in class \Class{Active|Resources} (in short, \Class{AR}) has a smell of Feature Envy that envies the class \Class{Resource|Weak|Reference} (in short, \Class{RWR}).
In addition, the class \Class{AR} has a smell of God Class.
The refactoring sequence recommended by MORCoRA includes two \Refactoring{Move Method} refactorings; to move the method \Method{cleanup|Active|Reference()} to class \Class{RWR}, and from the same source class, move the method \Method{deactivate()} to class \Class{Engine|Resource}.
These refactorings can improve the code quality and eliminate the code smells, and the semantic coherence of the source and target classes is also preserved.
Class \Class{AR} and \Class{RWR} are in the same file, and the developers having expertise on them are the same set, $\dev{A}$, $\dev{B}$, and $\dev{C}$.
The developers having expertise in class \Class{Engine|Resource} are $\dev{A}$ and $\dev{D}$.
Developer $\dev{A}$ participated in two in-progress issues, and he/she was too busy to review according to our settings.
For the remaining three developers, considering the expertise should cover all files related in the sequence, two candidate groups of reviewers $\dev{B}$ and $\dev{D}$, $\dev{C}$ and $\dev{D}$ can be selected.
MORCoRA recommended $\dev{B}$ and $\dev{D}$ to be the reviewers because this group had a higher $\RA$.
We checked the candidate groups' history contributions.
Developer $\dev{B}$ had two history commits, and $\dev{C}$ had only one.
The developer $\dev{B}$ added a Javadoc comment, applied refactorings, and added annotations for methods in the file, whereas developer $\dev{C}$ only changed the type of a field variable once, which indicates developer $\dev{B}$ having higher expertise than $\dev{C}$.
As a result, it is reasonable for MORCoRA to recommend this refactoring sequence and developers $\dev{B}$ and $\dev{D}$ to be the reviewers.
This solution explains how MORCoRA can successfully search refactoring sequences that conform to all three objectives.

\Conclusion{
The proposed method MORCoRA can effectively search for meaningful and reviewable refactoring sequences that can improve code quality.
We found that 58.3\% of solutions searched by MORCoRA have the ability to eliminate code smells while remaining meaningful and reviewable.
}

\subsection{\RQ{2}: \RQtwo} \label{ss:rq2}
\subsubsection{Motivation and Study Design}\label{sss:rq2_motivation}
Diverse search algorithms significantly impact the performance of search-based techniques.
To find the best search algorithm for MORCoRA to search refactoring sequences that conform to the three objectives, we set this RQ.
We compared NSGA-II with the other three evolutionary search algorithms: SPEA2~\cite{zitzler1998multiobjective}, IBEA~\cite{zitzlerIndicatorBasedSelectionMultiobjective2004}, MOCell~\cite{nebroMOCellCellularGenetic2009}, and a non-evolutionary search algorithm: Random Search (RS)~\cite{karnopp1963random} to prove the necessity of multi-objective evolutionary algorithms, following the guideline~\cite{arcuri2011practical}.
The outputs of all five algorithms were \pfs.
We performed five repeated runs for each of the algorithms because of the stochastic nature of the algorithms, and solutions with non-negative $\QG$ were selected from the \pfs of each algorithm.
The distributions of the objective scores of the selected solutions were compared to evaluate the performance of each algorithm.

\begin{figure}[tb]
  \centering
  \includegraphics[width=10.5cm]{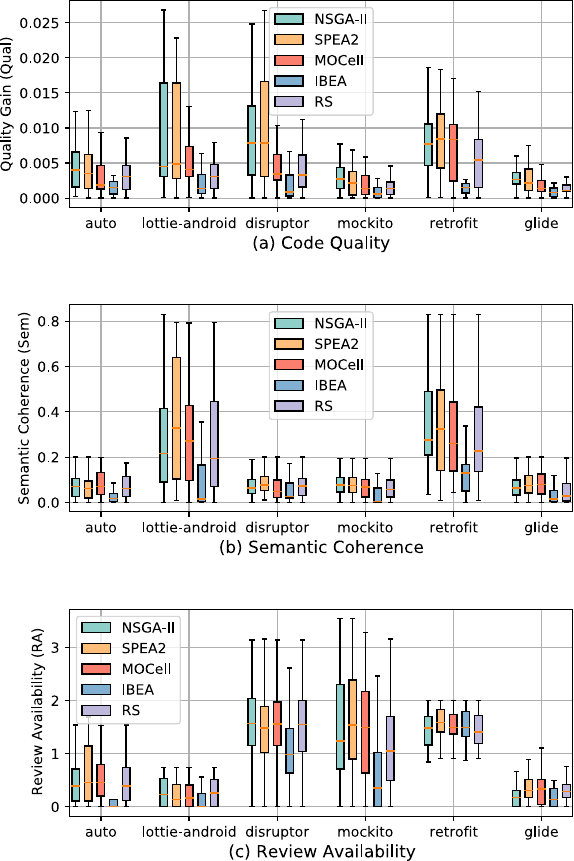}
  \caption{Distributions of the solutions for five algorithms on three objectives.}
  \label{f:rq2}
\end{figure}

\subsubsection{Results and Discussion}\label{rq1_results_and_discussion}
The objective scores are represented as boxplots, and their distributions are presented in \cref{f:rq2}.
Outliers are not shown in the figure because they only represent a few solutions, and our investigation target should be the distribution of the most solutions.

\Cref{f:rq2}(a) shows the distribution of solutions for the improving code quality objective. 
The y-axis is the improved $\QG$ after the refactorings are applied.
The scale for $\QG$ in different repositories varies 
because the optimal numbers of refactorings required by repositories are different.
The performance of different algorithms was compared by the median for boxes in each repository.
We found that NSGA-II had the best overall performance.
It outperformed at least two algorithms in all repositories by an average of 155.3\%.
It achieved the highest median in four out of six repositories and ranked second and third in the remaining two repositories.
Its solutions' average $\QG$ exceeded the RS algorithm by an average of 72.9\% in all repositories.
The algorithm SPEA2 had a similar performance to NSGA-II. 
It outperformed at least three other algorithms in all repositories and achieved the best in the two larger-size repositories \repository{lottie-android} and \repository{retrofit}.
The solutions of SPEA2 outperformed RS by an average of 66.5\%.
The algorithm MOCell achieved the third-highest median in four of six repositories and the second and fourth-highest in the remaining repositories.
The algorithm with the worst performance was IBEA. 
Its solutions achieved the lowest $\QG$ in all repositories.
The RS outperformed IBEA but was inferior to the other three MOEAs.

\Cref{f:rq2}(b) illustrates the distribution of solutions on the preserving semantic coherence objective, where the y-axis is the $\SC$.
We used the same method as in the objective Improving Code Quality; the comparison of different algorithms is through the median of each distribution.
SPEA2 demonstrated the best overall performance and had the highest $\SC$ in 50\% repositories.
In the repository~\repository{mockito}, NSGA-II performed best, and third in the remaining repositories.
Despite in the repository~\repository{disruptor} where RS ranked second, RS underperformed to NSGAII, SPEA2, and MOCell in all repositories.
IBEA performed the worst across all repositories.

\Cref{f:rq2}(c) represents the distribution of solutions on the review availability objective, where the y-axis is the $\RA$.
The median values of each distribution were used for comparison.
The NSGA-II outperformed at least two algorithms in the repositories~\repository{lottie-android}, \repository{disruptor}, and \repository{mockito}, and ranked fourth in the remaining three repositories.
In contrast, SPEA2 showed the opposite trend with NSGA-II.
It outperformed all other algorithms in~\repository{auto}, \repository{mockito}, and \repository{retrofit}.
MOCell showed the best overall performance.
It exceeds at least three algorithms in all repositories and achieved the highest median value in the repository~\repository{glide}.
Except in repository~\repository{retrofit}, the IBEA performed worst in all other repositories, and the median of it on repository~\repository{auto} and \repository{lottie-android} is zero.
Though RS showed the highest median value in repository~\repository{lottie-android}, it achieved low scores in others.

In general, NSGA-II exhibited a stronger performance than other algorithms though it did not perform the best across all repositories with respect to the three objectives.
The performance of SPEA2 was comparable to NSGA-II and sometimes even surpassed it.
However, the performance of SPEA2 on the objective of improving code quality was relatively poorer than that of NSGA-II.
The performance of the MOCell algorithm was not as good as NSGA-II or SPEA2, but it still outperformed RS.
The IBEA algorithm was ineffective in searching for solutions that conform to the three objectives. 
It showed poor performance on all three objectives, even worse than RS.
In the objectives of improving code quality and preserving semantic coherence, the RS only outperformed IBEA and was inferior to the other algorithms.
While it displayed slightly better performance in terms of possessing review availability, it still demonstrated inferior performance compared to NSGA-II, SPEA2, and MOCell.

One of the possible reasons that IBEA performed the worst is that it is highly sensitive to the parameter settings~\cite{arcuriParameterTuningSearch2011}.
The settings used in the experiment may not be the most appropriate for it, resulting it providing many duplicated non-dominated solutions in the search and failing to explore more solution space~\cite{xue2016ibed}.

\Conclusion{
We found that it is more effective to use evolutionary algorithms NSGA-II, SPEA2, and MOCell to search for refactoring sequences that conform to the three objectives than RS.
Among the four evolutionary algorithms used in the experiment, NSGA-II exhibited the best performance.
The algorithm SPEA2 showed slightly inferior performance than NSGA-II.
The algorithm MOCell performed better than RS in most of the cases, but not as well as NSGA-II or SPEA2.
The algorithm IBEA was ineffective in searching for solutions that conform to the three objectives under the settings of this paper.
It underperformed RS in most of the situations.
}

\subsection{\RQ{3}: \RQthree} \label{ss:rq3}
\subsubsection{Motivation and Study Design}\label{sss:rq3_motivation}
To show the effectiveness of the newly introduced objective \textit{possessing review availability}, we investigate its impact to explore the effectiveness of the designed metric.
By quantitative analysis, the impact can be observed through design metric values.
Through qualitative analysis, the impact is further assessed, and the reason for the impact is also discussed.

\begin{figure}[t]\centering
  \includegraphics[width=\textwidth]{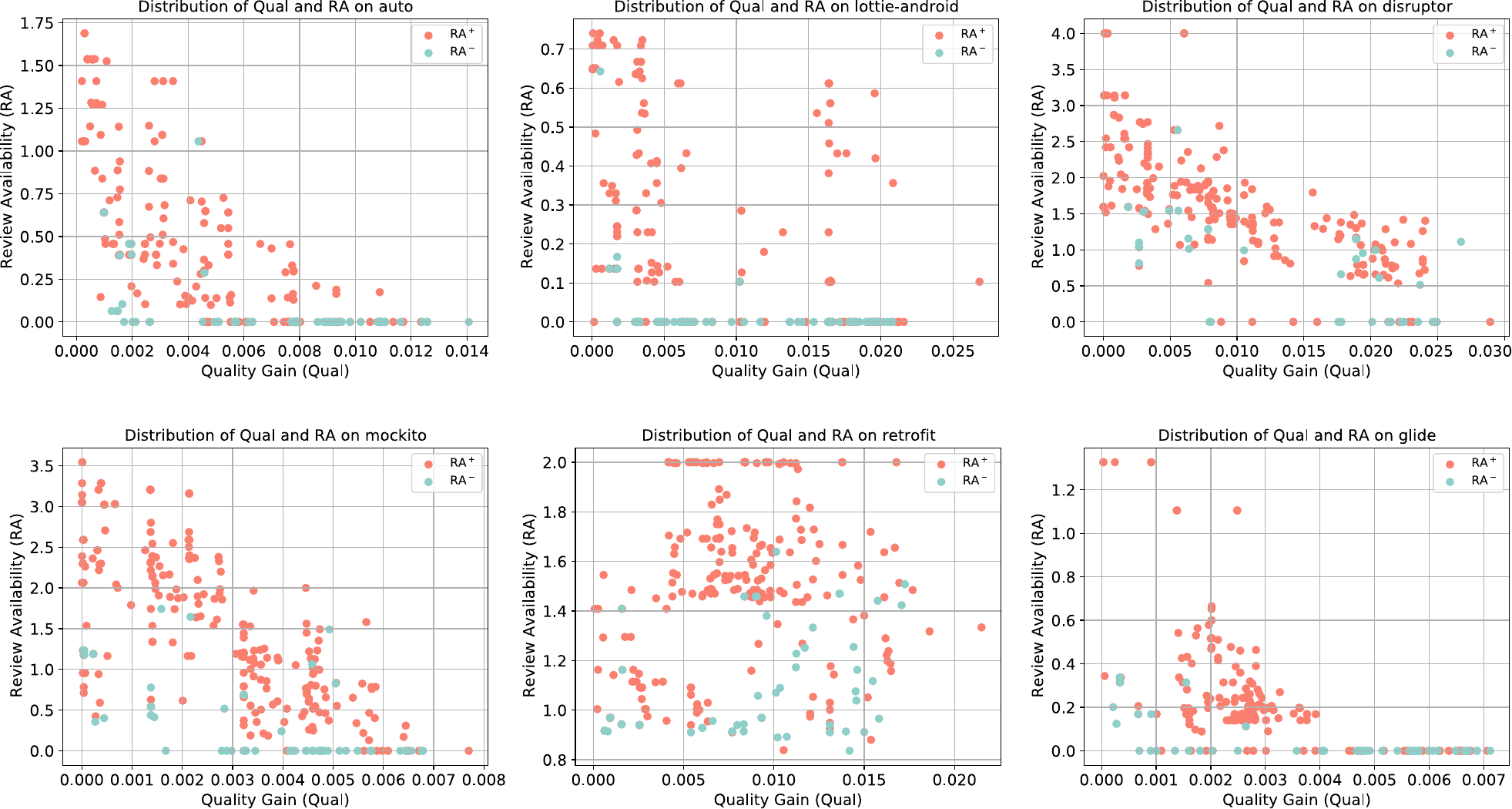}
  \caption{Distributions of the solutions on $\QG$ and $\RA$.}
  \label{f:rq3}
\end{figure}

\begin{figure}[t]\centering
  \includegraphics[width=\textwidth]{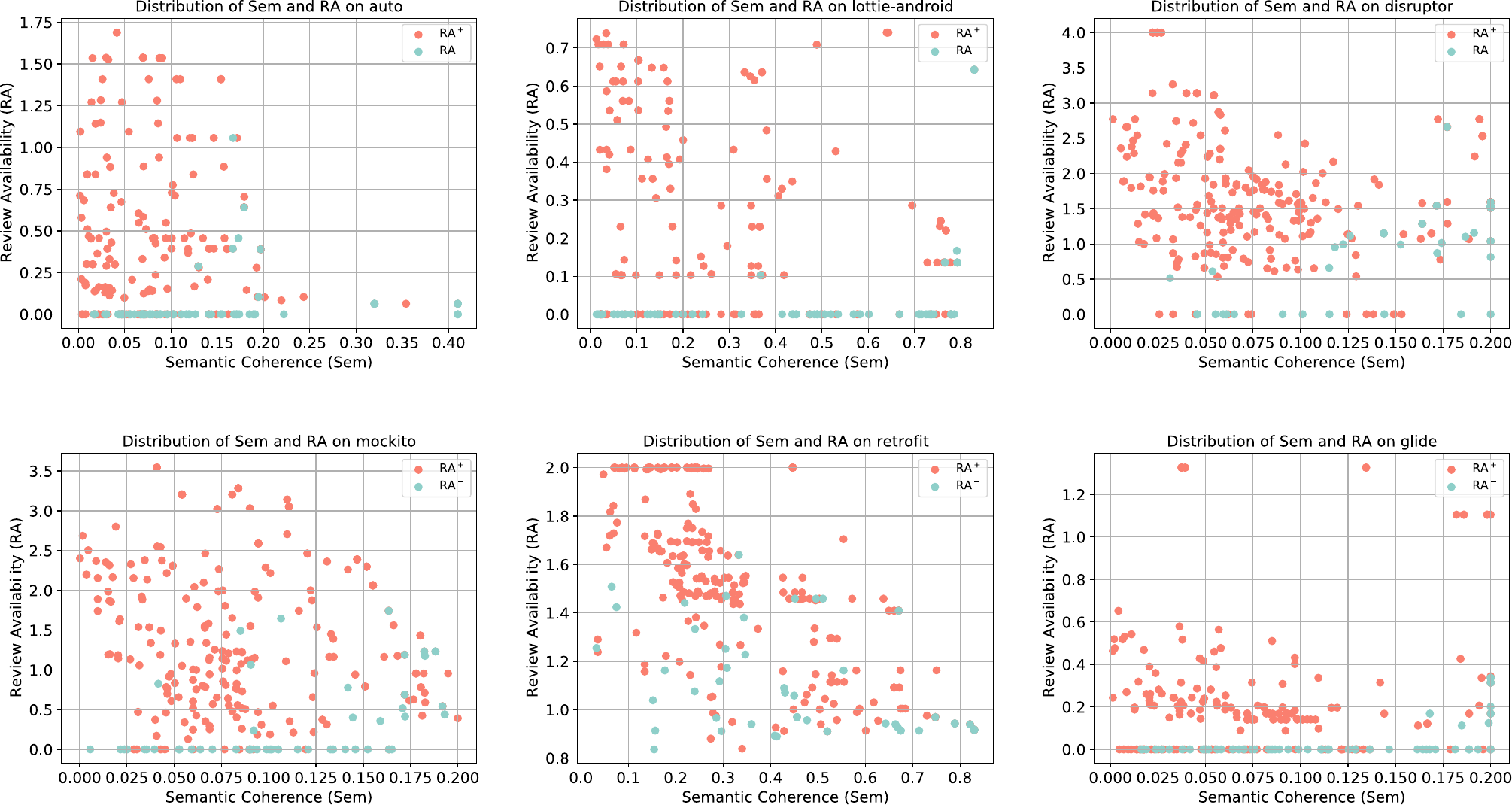}
  \caption{Distributions of the solutions on objectives $\SC$ and $\RA$.}
  \label{f:rq3_sem}
\end{figure}

To answer this question, we compared MORCoRA, denoted as $\RAY$ (with the review availability objective), with the technique proposed by Ouni et al.~\cite{ouni2012search}, denoted as $\RAN$ (without the review availability objective), which considers only two objectives: improving code quality and preserving semantic coherence.
Since there is no publicly available implementation for the baseline technique by Ouni et al., we reproduced it according to their paper for comparison.
The solutions derived from $\RAY$ and $\RAN$ were compared and analyzed from quantitative and qualitative perspectives.
In quantitative analysis, we compared distributions of solutions to investigate the improvement of $\RA$ by considering the possessing review availability objective.
In qualitative analysis, we examined the refactoring sequences recommended by $\RAY$ and $\RAN$ to better explain the merit brought by $\RAY$.
By finding solutions that can eliminate code smells and are meaningful but not reviewable, and those solutions are only recommended by $\RAN$ but not by $\RAY$, we are able to prove the usefulness of $\RAY$.
The tool Jdeodorant is used to identify code smells.
Because solutions with negative $\QG$ are meaningless refactoring sequences, only solutions with positive $\QG$ derived from $\RAY$ and $\RAN$ were selected for comparison.

\subsubsection{Results and Discussion for Quantitative Analysis}\label{rq2_results_and_discussion_for_quantitative_analysis}
The scatter plots, which show the distribution of solutions from $\RAY$ and $\RAN$ on objectives improving the code quality and possessing review availability, preserving semantic coherence and possessing review availability are represented in \cref{f:rq3,f:rq3_sem}, respectively.
Each plot represents the distribution for one repository. 
The x-axis in \cref{f:rq3} represents the $\QG$ value and that in \cref{f:rq3_sem} does the $\SC$ value, and the y-axis shows the $\RA$ value.
Each dot in the figures represents the $\QG$ and $\RA$ value or the $\SC$ and $\RA$ of one refactoring sequence.
Note that we found the distributions of $\SC$ and $\RA$, $\QG$ and $\RA$ produced similar results, we only discuss the result of $\QG$ and $\RA$ below.

It can be observed that the number of data points for $\RA^+$ is greater than $\RA^-$ in all repositories.
This is caused by the adding of an objective will enlarge the size of the  \pfs of searched results, which leads to $\RA^+$ having 3.99 times more points than $\RA^-$ averagely in \cref{f:rq3}.

For all plots, the data points for $\RA^+$ tend to be positioned at higher locations than the data points for $\RA^-$, which indicates that $\RAY$ performed superior to $\RAN$ on the objective of possessing high review availability by having higher $\RA$ for all repositories.
We can also observe that solutions with the same $\QG$ but higher $\RA$ existed in all repositories, which indicates that $\RA^+$ has the ability to recommend solutions with acceptable $\QG$ and $\SC$ but better $\RA$ than $\RA^-$.

\begin{table}
    \centering
    \tbl{Reviewable Solution Ratio\label{t:reviewable_ratio}}
    {\begin{tabular}{@{}lccc@{}}
          \toprule
          Repository & $\RAY$ & $\RAN$ & Relative improvement\\
          \midrule
            \repository{auto} & 0.77 & 0.19 & 4.05\\
            \repository{lottie-android} & 0.73 & 0.14 & 5.21\\
            \repository{disruptor} & 0.93 & 0.81 & 1.15\\
            \repository{mockito} & 0.95 & 0.68 & 1.40\\
            \repository{retrofit} & 1.00 & 1.00 & 1.00\\
            \repository{glide} & 0.73 & 0.16 & 4.56\\
        \midrule
        Average (excluding \repository{retrofit}) & & & 3.27\\
        \botrule
    \end{tabular}}
\end{table}

In addition, solutions with both higher $\QG$ and $\RA$ were found in repositories \repository{lottie-android} and \repository{retrofit}.
Some modules, contributed by a certain group of developers, tend to need more refactorings than others.
Considering the objective of review availability can help the search algorithm focus on these modules, leading to higher $\QG$ and $\RA$ solutions.

The zero $\RA$ indicates that the solution can be regarded as not reviewable because no reviewer fits the requirement of enough expertise and a low workload.
The reviewable solution ratios for each repository of both $\RAY$ and $\RAN$ are shown in \cref{t:reviewable_ratio}.
It can be observed that except the repository \repository{retrofit}, $\RAY$ had an average of 3.27 times more reviewable solutions than $\RAN$.
Note that the reason that all solutions in \repository{retrofit} were reviewable was caused by the existence of a \textit{common reviewer} in it.
The \textit{common reviewer} had expertise on 99.3\% of files, and he/she was regarded as free of workload in the time period in the experiment setting.
Thus, he/she could be regarded as the reviewer for all the solutions in both $\RAY$ and $\RAN$.

The positive $\RA$ represents the expertise of the proper reviewers own on the solutions.
For all solutions in the six repositories, although $\RA^+$ has an average $\QG$ value of only 75.2\% of $\RA^-$ and an average $\SC$ value of 67.0\% of $\RA^-$, its average $\RA$ value is 533.8\% of $\RA^-$.
This indicates that $\RA^+$ yields solutions of slightly lower but acceptable $\QG$ and significantly higher $\RA$.

\Conclusion{%
The $\RA^+$ can effectively search solutions with acceptable $\QG$ and higher $\RA$ than the existing search-based refactoring technique.
The solutions recommended by $\RA^+$ on our dataset had an average of 24.8\% lower $\QG$ and 33.0\% lower $\SC$ but 433.8\% higher $\RA$ than the $\RA^-$.
}

\subsubsection{Results and Discussion for Qualitative Analysis}\label{rq2_results_and_discussion_for_qualitative_analysis}
Some refactoring sequences have zero $\RA$ values and are 
 recommended only by $\RA^-$ but not by $\RA^+$. 
 Two of them are shown below, that they can improve the code quality and mitigate code smells but were not review-available in practice.

The class \Class{Auto|Builder|Processor} (in short, \Class{ABP}) in the repository \repository{auto} has a smell of God Class, and a refactoring sequence containing one refactoring was recommended by $\RA^-$ to mitigate the code smell.
The refactoring is to move a member field \Field{java|Lang|Void} of class \Class{ABP} to class \Class{Qualified|Dependency|Impl}.
The developers regarded as with expertise in the source class are $\dev{A}$ and $\dev{B}$, whereas developers $\dev{A}$, $\dev{C}$, and $\dev{D}$ have expertise on the target class.
Based on their workload, only $\dev{B}$ is available in the time period used in the experiment settings.
As a result, no proper reviewers can be selected for this refactoring.
The reviewer $\dev{A}$ has expertise in both source and target class, but his/her in-progress workload was high; he/she participated in a total of 13 pull requests and issues during the time period.
The reviewer $\dev{C}$ and $\dev{D}$ only had experience with the target classes, and both of them were working on two other pull requests. 

In the same repository, another refactoring sequence with two refactorings was recommended.
The class \Class{Immutable|Map|Serializer} has a code smell of God Class, and the two refactorings are to move two methods \Method{generate|Key|Map|Function()} and \Method{generate|Value|Map|Function()} from it to another class \Class{Immutable|Map|Serializer|Extension}.
Only developer $\dev{A}$ possesses expertise in both source and target class, and he/she was engaged with 13 pull requests and issues.
Consequently, this refactoring sequence is unreviewable.

From the two solutions above, we can conclude that MORCoRA avoids recommending refactorings, which will be applied to files where all reviewers with expertise are having a heavy workload.
Those refactorings cannot be executed due to the absence of qualified reviewers; therefore, it is crucial to consider the developers' buffer when evaluating recommended refactorings.

\Conclusion{%
By manually assessing and comparing the refactoring sequences searched by MORCoRA with the existing technique, we confirmed that the MORCoRA can avoid recommending refactorings with low review availability.
Additionally, we emphasize the importance of considering the developers' buffer when evaluating such recommended refactorings.
}

\section{Threats to Validity}\label{s:threats}
In this section, we discuss four types of potential threats to the validity of our work as follows.

\subsection{Internal Validity}\label{ss:internal_validity}
One of the possible threats to internal validity is that the contributors may use different aliases in Git and on GitHub, which may affect the calculation of reviewers' expertise and workload.
We mitigated it by manually identifying and unifying the aliases based on the information collected from the contributors' GitHub Personal page and their self-introduction webpages.
However, despite these efforts, there remains a risk that some aliases were not captured or correctly linked.

Another threat is that using participation in pull requests and issues may not necessarily mean that the participants are busy and unable to review.
However, it has been proven by Chouchen et al.~\cite{chouchen2021whoreview} that participation in pull requests and issues can be used to evaluate workload effectively.
Thus, we still adopted it into our work.

Additionally, there is a potential threat related to the manual assessment in \cref{ss:rq1}, where 60 solutions were evaluated to demonstrate the effectiveness of the proposed technique.
Though we carefully conducted the manual assessment process to ensure its validity, we cannot fully guarantee its effectiveness, as we are not the contributors to the target projects. 
The solutions and assessment results are included in the replication package~\cite{dataset}, and researchers are encouraged to investigate the validity of our findings.

\subsection{External Validity}\label{ss:external_validity}
Although we selected six popular open-source repositories with considerable diversity in size, number of commits, and number of pull requests and issues, we cannot claim that the same results can be generalized to other repositories.
By substituting JxPlatform2 with alternative static analysis tools, our techniques can be adapted to support other object-oriented languages.
In addition, the qualitative analysis was conducted by us, which may be biased.

\subsection{Construct Validity}\label{ss:construct_validity}
In this study, we used repositories collected from GitHub and used the GitHub pull requests and issues history as the measurement of reviewers' workload.
In other words, the observation of reviewers' workload was only from the records in GitHub.
However, reviewers may have other workloads, such as participating in other projects or non-programming workloads, which are not considered in this study.
To solve the above problem, a study about inquiring reviewers' workload is necessary, which is one of the future work.

\subsection{Conclusion Validity}\label{ss:conclusion_validity}
We cannot claim that the hyperparameter settings are optimal, which could adversely affect the experimental results.

\section{Conclusion and Future Work}\label{s:conclusion}
In this study, we proposed MORCoRA, a multi-objective refactoring recommendation tool recommending refactorings that conform to three objectives: 1) improving code quality, 2) preserving semantic coherence, and 3) possessing high review availability.
We performed experiments by using the technique on six open-source Java repositories to evaluate it.
The results of quantitative and qualitative analyses confirmed that MORCoRA could effectively recommend refactoring sequences that possess high review availability.
We suggest that review availability should be considered in search-based refactoring studies and refactoring recommenders should consider both the impact on quality improvement and the developer resources required for review when recommending refactorings.

In this work, we solely focused on analyzing the workload of developers within the target project.
However, it is common for developers to contribute to multiple projects simultaneously during a given time period. To obtain a more accurate assessment of their workload, we intend to track their activity across all relevant repositories in the future.
In addition, the non-programming workload should also be taken into consideration.
Furthermore, we plan to invite participants from both academic and industrial fields to use MORCoRA and obtain their feedback to further improve it.

The replication package, which includes all refactoring sequences searched in the experiment of this study and the manually assessed results, is available~\cite{dataset}.

\subsection*{Acknowledgments}
This study was partly supported by JST SPRING No.\ JPMJSP2106, JSPS Grants-in-Aid for Scientific Research Nos.\ JP21H04877, JP21K18302, JP21KK0179, JP23K24823, and JP24H00692.


\end{document}